\documentclass[preprint]{aastex}
\usepackage{colordvi}
\usepackage{color}
\usepackage[hyphens]{url}

\begin{document}

\title{Relationship between Intensity of White-Light Flares and Proton Flux of Solar Energetic Particles}

\author{Nengyi Huang\altaffilmark{1,2},
        Yan Xu\altaffilmark{1,2},
        and Haimin Wang\altaffilmark{1,2}}

\affil{1.\ Space Weather Research Lab, Center for
Solar-Terrestrial Research, \\ New Jersey Institute of Technology \\
323 Martin Luther King Blvd, Newark, NJ 07102-1982}

\affil{2.\ Big Bear Solar Observatory, \\ New Jersey Institute of Technology \\
40386 North Shore Lane, Big Bear City, CA 92314}


Solar Energetic Particles (SEPs; see review by Reames 1999) are considered to be one of the most important kinds of events in terms of their effects on space weather. They are mostly in the form of as accelerated protons and heavy ions. The SEPs were found to be accelerated by magnetic reconnection during solar flares \citep[e.g.][]{Mori1998, Bombardieri2008}. They also can be accelerated by shocks while CMEs are propagating in interplanetary space \citep[e.g.][]{Klein2001, Roussev2004}. \citet{Aschwanden2012} found that both flare and CME shock plays important roles. In the solar observations,  white light flares (WLF) are considered as most energetic signature of particles bombarding solar surface and associated with both hard X-ray (HXR) and Gamma ray emissions \citep{Hudson2016}.  In our previous study, we found a close correlation between WLF intensities and the HXR power index \citep{Huang2016}. In this study, we focus on the comparison between SEP flux and WLF intensities.

According to the NOAA space environment services center, there were 43 SEP events recorded from 2010, when SDO was launched \citep{SDO}, to September 2017. The counts of SEP flux can be found at \url{https://umbra.nascom.nasa.gov/SEP/}. In this study, WLFs are identified using the 45 s WL data obtained by SDO/HMI \citep{SDO, HMI}. The WL emission was visually detected in 15 flares above M5. The WL emission is characterized by equivalent area \citep[EA,][]{Wang2008,Huang2016}, which is the integrated enhancement of contrast over the entire flare ribbons.

\begin{figure}[h!]
\centering
\includegraphics[scale=1]{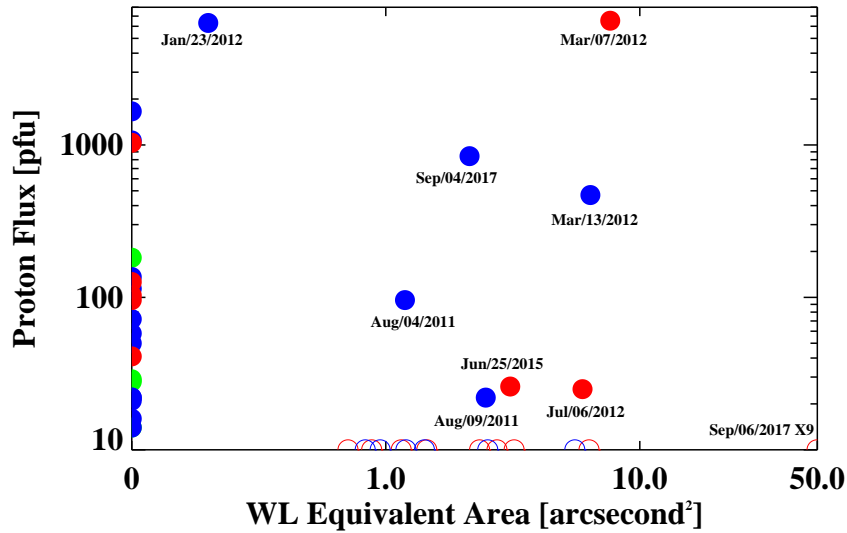}
\caption{Comparison of SEP events and WLFs. Solid circles represent SEP events and empty circles represent WLFs, which were not associated with SEP events. Different colors indicate the magnitudes of flares: red for X-class, blue for M-class and green for C-class flares.
\label{f1}}
\end{figure}

By comparing the SEP and WLF lists, the events can be divided into three groups as shown in Figure~\ref{f1}. The 1st group includes SEP events that were not associated with WLFs. They are plotted using solid circles (half) on y-axis indicating zero WL emission. Their X-ray flux (represented by GOES classes in different colors) does not correlate with the proton flux. As we can see, a C-class flare may have stronger proton flux than many X- and M-class flares. The 2nd group contains WLFs without SEP detected. Those events are represented by empty circles (half) on x-axis. The events in these two groups show that there is no correlation between WLFs and SEP in 39 out of 47 events. This result indicates that the SEPs may not be accelerated in the region where flare-related magnetic reconnection takes place. In the 3rd group, we found eight SEP events, which were associated with WLFs, including 5 M-class (blue color) and 3 X-class (red color) flares. As we can see in the plot, there is no clear correlation between SEP flux and EA of WLFs. For instance, the flare on Sep. 6, 2017 was the strongest WLF with EA almost an order of magnitude larger than the second largest one. However, no obvious increase of proton flux was detected. The proton fluxes were already enhanced due to the SEP event on Sep. 5, 2017. There was no significant continued enhancement of proton flux associated this X9 flare on Sep. 6, 2017.
On the other hand, another X-class flare on Mar. 7, 2012 has strong WL emission and is associated with the strongest SEP event.

In summary, our preliminary results show that most ($>83\%$) of WLFs and SEP events have no correspondence. Note that all SEP events were checked and some limb WLFs were excluded to avoid projection effect in calculating the EA. Therefore, the actual percentage of non-correlated events should be even higher. In a small group of events with both WL emission and SEPs, we did not see a positive correlation between SEP flux and contrast enhancement in WL. A straightforward speculation is that the acceleration process could be different for SEPs and the energetic electrons powering WLFs in the events analyzed.

SEP data are courtesy of NOAA Space Weather Prediction Center and SDO/HMI data are courtesy of NASA/SDO and HMI science team.

\bibliographystyle{aasjournal}

\end{document}